# Hybridization and Correlation between *f*- and *d*-orbital electrons in a valence fluctuating compound EuNi$_2$P$_2$

Z. X. Yin[1*], X. Du[1*], W. Z. Cao[2], J. Jiang[2,3], C. Chen[2,3], S. R. Duan[1], J. S. Zhou[1], X. Gu[1], R. Z. Xu[1], Q. Q. Zhang[1], W. X. Zhao[1], Y. D. Li[1], Yi-feng Yang[4], H. F. Yang[2,5], A. J. Liang[2,5], Z. K. Liu[2,5], H. Yao[6], Y. P. Qi[2,5], Y. L. Chen[1,2,5,7†], and L. X. Yang[1,8†]

[1]*State Key Laboratory of Low Dimensional Quantum Physics, Department of Physics, Tsinghua University, Beijing 100084, China*

[2]*School of Physical Science and Technology, ShanghaiTech University and CAS-Shanghai Science Research Center, Shanghai 201210, China.*

[3]*Advanced Light Source, Lawrence Berkeley National Laboratory, Berkeley, California 94720, USA*

[4]*Beijing National Laboratory for Condensed Matter Physics and Institute of Physics, Chinese Academy of Sciences, Beijing 100190, China*

[5]*ShanghaiTech Laboratory for Topological Physics, Shanghai 200031, China.*

[6]*Institute for Advanced Study, Tsinghua University, Beijing 100084, China*

[7]*Department of Physics, Clarendon Laboratory, University of Oxford Parks Road, Oxford, OX1 3PU, UK*

[8]*Frontier Science Center for Quantum Information, Beijing 100084, China.*

*These authors contributed equally to this work.

[†]*Email address: LXY: lxyang@tsinghua.edu.cn, YLC: yulin.chen@physics.ox.ac.uk*

**The interaction between localized *f* and itinerant conduction electrons is crucial in the electronic properties of heavy fermion and valence fluctuating compounds. Using high-resolution angle-resolved photoemission spectroscopy, we systematically investigate the electronic structure of the archetypical valence fluctuating compound EuNi$_2$P$_2$ that hosts multiple *f* electrons. At low temperatures, we reveal the hybridization between Eu 4*f* and Ni 3*d* states, which contributes to the electron mass enhancement, consistent with the periodic Anderson model. With increasing temperature, interestingly, we observe opposite temperature evolution of electron spectral function above and below the Kondo coherence temperature near 110 K, which is in contrast to the monotonic valence change and beyond the expectation of the periodic Anderson model. We argue that both *f-d* hybridization and correlation are imperative in the electronic properties of EuNi$_2$P$_2$. Our results shed light on the understanding of novel properties, such as heavy fermion behaviors and valence fluctuation, of rare-earth transition-metal intermetallic compounds with multiple *f* electrons.**

Rare-earth transition-metal intermetallic compounds (RTICs) provide a fertile playground to explore various fascinating emergent properties with strong electronic correlation [1-5]. One prominent example is the valence fluctuation effect, where *f*-electron states are partially occupied, inducing the non-integral or intermediate valency of rare-earth elements [6,7]. Historically, two paradigmatic theoretical frameworks have been established to understand this mysterious effect. The first one interprets the valence fluctuation as a quantum mechanical mixing of two integral valence states induced by the hybridization between localized 4*f* and itinerant conduction electrons, as described by the periodic Anderson model (PAM), which is an effective model in the study of heavy fermion systems [8,9]. The other theory, by contrast, considers the valence fluctuation as arising from thermal fluctuation between two different integral valences induced by Coulomb interaction between 4*f* and conduction electrons (FK interaction), as proposed in the Falicov-Kimball model (FKM) [10,11].

The situation is, however, much more complicated in realistic materials due to the entanglement of the hybridization and Coulomb interaction between *f* and conduction electrons. While Ce- and Yb-based valence fluctuating systems are generally well described by the PAM [12-14], the FK interaction is proposed to be important for understanding the valence transition in $YbInCu_4$ and γ-α transition of Ce metal [15,16]. Moreover, the valence fluctuation derived from FK interaction serves as a pairing mechanism for the enhanced superconductivity in heavy fermion system $CeCu_2Si_2$ [17] and induces the non-Fermi liquid behavior near the quantum critical point of the Doniach phase diagram [17,18], which strongly challenges the conventional spin fluctuation mechanism. Therefore, it is crucial to investigate the hybridization and correlation between *f* and conduction electrons to understand novel properties of RTICs, such as heavy fermion behavior and valence fluctuation effect. Particularly, it is highly desirable to experimentally investigate RTICs with multiple *f* electrons that are rarely explored.

In this letter, using high-resolution angle-resolved photoemission spectroscopy (ARPES) and *ab-initio* calculation, we comprehensively study the electronic structure of a prototypical valence fluctuating material EuNi$_2$P$_2$ with multiple *f* electrons [19]. At low temperatures, we observe the hybridization between Eu 4*f*- and Ni 3*d*-states, as manifested by the gap opening near *f-d* band crossings and a kink near the Fermi level ($E_F$), reminiscent of Ce-based heavy fermion systems and dynamical mean-field theory (DMFT) calculations of PAM [20,21]. The *f-d* hybridization contributes to electron mass enhancement at low temperatures, which explains the moderate heavy-fermion behavior of EuNi$_2$P$_2$. In contrast to the monotonic valence change with temperature, both the spectral functions of Eu 4*f*- and Ni 3*d*-electrons show opposite temperature evolution below and above the Kondo coherence temperature near 110 K, where the resistivity curve shows a local maximum [22]. We argue that the FK interaction should be considered in understanding the valence fluctuation effect in Eu-based compounds and, possibly, in other *f*-electron systems.

High-quality single crystals of EuNi$_2$P$_2$ were synthesized by a Sn-flux method [22]. High-resolution ARPES measurements were performed at beamline 4.5.1 of Stanford Synchrotron Radiation Light Source (SSRL), beamline SIS of Swiss Light Source (SLS), and beamline BL03U of Shanghai Synchrotron Radiation Facility (SSRF). Data were collected with Scienta R4000 (DA30) electron analyzers at SSRL and SLS (SSRF). The overall energy and angular resolutions were set to 15 meV and 0.2°, respectively. The samples were cleaved *in situ* and measured under ultra-high vacuum less than $1.0 \times 10^{-10}$ mbar. First-principles band structure calculations were performed using QUANTUM ESPRESSO code package [23] with a plane wave basis. The exchange-correlation energy was considered under Perdew-Burke-Ernzerhof

(PBE) type generalized gradient approximation (GGA) [24]. The *f* electrons of Eu atoms were treated as core states. The cutoff energy for the plane-wave basis was set to 480 and 960 eV for calculation without and with spin-orbit coupling, respectively. A Γ-centered Monkhorst-Pack k-point mesh of 11 ×11 × 5 was adopted for a self-consistent charge density.

EuNi$_2$P$_2$ crystallizes in a body-centered tetragonal ThCr$_2$Si$_2$- (I-) type structure with two Ni-P layers separated by a Eu layer in each unit cell [Fig. 1(a)]. The electrical resistivity shows a local maximum near 110 K [Fig. 1(b)], consistent with previous reports [22] and similar to Ce-based heavy fermion materials [2,25,26]. While this local maximum is usually attributed to the Kondo coherence temperature in heavy fermion systems, it can also be derived from the renormalization of density of states near $E_F$ in FKM [27]. The temperature-dependent magnetic susceptibility in Fig. 1(b) exhibits a Curie-Weiss behavior at high temperatures and saturates below 50 K, which is commonly observed in valence fluctuating materials and can be explained either by the phenomenological inter-configuration fluctuation (ICF) model [19] or the coherence of Kondo resonance [22]. Fig. 1(c) shows the X-ray photoemission spectrum with the characteristic Eu, Ni and P core-level peaks. We observe both Eu$^{2+}$ and Eu$^{3+}$ 3*d* peaks, confirming the valence fluctuating nature of EuNi$_2$P$_2$ [28].

Figure 2 presents the electronic structure of EuNi$_2$P$_2$. On the Fermi surface (FS), we observe a large four-point-star like hole pocket around the $\bar{\Gamma}$ point, which crosses the Brillouin zone (BZ) edge [marked as α, the black dashed curve in Fig. 2(a)]. In addition, there exists another hole pocket β around the $\bar{\Gamma}$ point in the second BZ, suggesting a $k_z$ dependence of the Fermi surface. Our *ab-initio* calculation in Figs. 2(b) and 2(c) captures the general FS structure, including the large and small hole pockets around $\bar{\Gamma}$ [Fig. 2(c)]. It is worth noting that the FS of EuNi$_2$P$_2$ is mainly constructed by Ni 3$d$ orbitals (Supplemental Material, Fig. S2 [29]). Figures 2(d)-(g) show the band structure along high-symmetry directions. Consistent with the FS structure, we observe an electron-like band α (from the large hole pocket in the second BZ) and a hole-like band β around $\bar{\Gamma}$, which are better visualized in the second BZ [Figs. 2(d)-(f)]. The measured band structure is nicely reproduced by *ab-initio* calculation of Ni 3$d$ bands after renormalized by a factor of about 1.25 [the dashed lines in Figs. 2(d)-(g)], suggesting a weak electronic correlation effect of Ni 3$d$ electrons. In addition to the α and β bands that cross $E_F$, Ni 3$d$ orbitals also form dispersive band γ at high binding energies below 1 eV.

Prominently, we observe multiple flat bands between -0.7 eV and $E_F$, which intersect with the dispersive conduction bands [Figs. 2(d)-(g)]. These multiplets are identified as Eu 4$f^6$ final states carrying different total angular momenta [30,31]. Previous ARPES measurements have shown the hybridization between Eu 4$f$ flat bands and Ni 3d valence band that is far below $E_F$ [the γ band in Fig. 2(g)] at specific momentum region [31]. Figure 3 investigates the hybridization between Eu 4$f$ multiplets and the conduction bands that cross $E_F$, i.e., the α and β bands. In Fig. 3(a), we notice that the flat band F$_6$ near -0.75 eV shows a clear curvature near the $\bar{\Gamma}$ point as indicated by the green arrow, suggesting the hybridization with the α band. The hybridization between the flat $f$ bands and α band is better visualized along $\bar{X}\bar{M}$, which opens multiple gaps, as indicated by the blue arrows in Figs. 3(b) and 3(c). Interestingly, we observe a kink-like structure in the α band along $\bar{X}\bar{M}$ near -40 meV, as shown by the red arrow in the zoom-in plot of ARPES image and momentum distribution curves (MDCs) in Figs. 3(d) and 3(e). By fitting the MDCs to a Lorentzian, the kink is clearly observed in the extracted α band dispersion [Fig. 3(f)]. Since the $F_0$ band resides at about -0.1 eV, we emphasize that the kink is not induced by the hybridization between the α band and the $F_0$ band. Instead, it is reminiscent of the kink induced by Kondo resonance in Ce-based heavy fermion systems [20,21,32,33] and can be well understood by the PAM, which considers the interaction between the Kondo resonance and the conduction band. With the renormalized *f* level (the Kondo resonance) energy $\varepsilon_0$ and renormalized hybridization energy $V_k$ as fitting parameters, the fit to PAM nicely reproduces the observed kink, giving $\varepsilon_0$=24 meV and $V_k$=56 ± 5 meV [orange curve in Fig. 3(f), Supplemental Material, Fig. S3 [29]], similar to the value of $V_k$ in CeCoGe$_{1.2}$Si$_{0.8}$[32]. The observed *f-d* hybridization, according to the PAM, can induce the heaviness of electrons. Indeed, the Fermi velocity of the α band is reduced by a factor of about 2.2, suggesting an enhancement of effective electron mass [22,34,35]. Although this value is much smaller than that obtained from specific heat measurement, it is consistent with the fact that ARPES usually reports a much smaller mass enhancement than that in transport measurements [32,36].

To further investigate the interplay between localized 4*f* electrons and itinerant 3*d* electrons, we conduct temperature-dependent ARPES measurements up to 205 K in Figure 4. Fig. 4(a) presents the temperature evolution of MDCs at $E_F$ along $\bar{X}\bar{M}$ to track the shift of the α band, and Fig. 4(b) presents the temperature evolution of the energy distribution curves (EDCs) integrated over the momentum window of [0.4, 0.7] Å$^{-1}$ along $\bar{\Gamma}\bar{X}$, where only flat-bands contribute to the photoemission spectra (Supplemental Materials, Figs. S4-S6 [29]). Unexpectedly, we extract a non-monotonic shift of both the α band and the flat bands above and below 110 K, as summarized in Fig. 4(c). We also observe a non-monotonic change of the spectral weight of the flat bands with temperature, as shown in Figs. 4(d) and 4(e). Both the band shift and the change of flat-band spectral weight are reminiscent of the resistivity of EuNi$_2$P$_2$, suggesting a correlation between the evolution of the electronic structure and transport properties.

The spectral evolution below 110 K is consistent with the valence change from Eu$^{2+}$ to Eu$^{3+}$ at low temperatures. However, the spectral response above 110 K, i.e., the reversed shift of both $f$ and conduction bands, is in drastic contrast to the monotonic valence change with the temperature and cannot be understood within the classical PAM. Instead, previous optical spectra suggest that the correlation between the localized $f$ and itinerant conduction electrons (FK interaction), as in the FKM, should be considered [34]. Similar to the correlation effect in the Hubbard model, the FK interaction also tends to split the non-interacting spectral function of electrons into lower and upper sub-bands, only this Hubbard-like splitting occurs for both $f$ and conduction electrons [27,37,38]. Moreover, the FK interaction can be enhanced by increased $f$ electron concentration at high temperatures, similar to the situation in YbInCu$_4$ and EuNi$_2$Si$_{1-x}$Ge$_x$ [39], which induces the reversed shift of both $f$ and conduction bands and spectral weight transfer from lower to higher sub-bands above 110 K [27,39,40]. The relevance of the FK interaction is further supported by the observation of an energy gap at the Fermi momentum of the α band above 110 K as shown in Figs. 4(f) and 4(g) [27,39].

Therefore, our results suggest that on the one hand, spin and charge fluctuation play an important role in the low-temperature properties of EuNi$_2$P$_2$ according to the PAM, as suggested by the direct observation of the $f$-$d$ hybridization and the band renormalization by Kondo resonance [41]. On the other hand, the FK interaction may further enhance the spin and charge fluctuations [42-44]. The synergetic effect of hybridization and correlation between the *f* and conduction electrons makes EuNi$_2$P$_2$ a unique system among Eu-based valence fluctuating materials [7], which exhibits not only a heavy fermion state originated from Kondo effect [22,45], but also a genuine valence fluctuation behavior with a Eu-valence of 2.5 at the ground state [19].

In conclusion, we systematically investigate the electronic structure of EuNi$_2$P$_2$ and its temperature evolution using ARPES. We suggest that both the hybridization and correlation effects between the *f* and conduction electrons should be considered to fully understand the electronic properties of EuNi$_2$P$_2$. Our results provide important insights into the microscopic interaction between the localized *f* and itinerant conduction electrons in a multi-*f*-electron system other than the widely investigated Ce- and Yb-based compounds, which will shed light on the understanding of the intriguing properties of RITCs such as heavy fermion and valence fluctuation effects.

**Acknowledgment**

This work was supported by the National Natural Science Foundation of China (Grants No. 11774190, No. 116427903, No. 11634009), the National KeyR&D program of China (Grants No. 2017YFA0304600 and No. 2017YFA0305400), EPSRC Platform Grant (Grant No. EP/M020517/1).

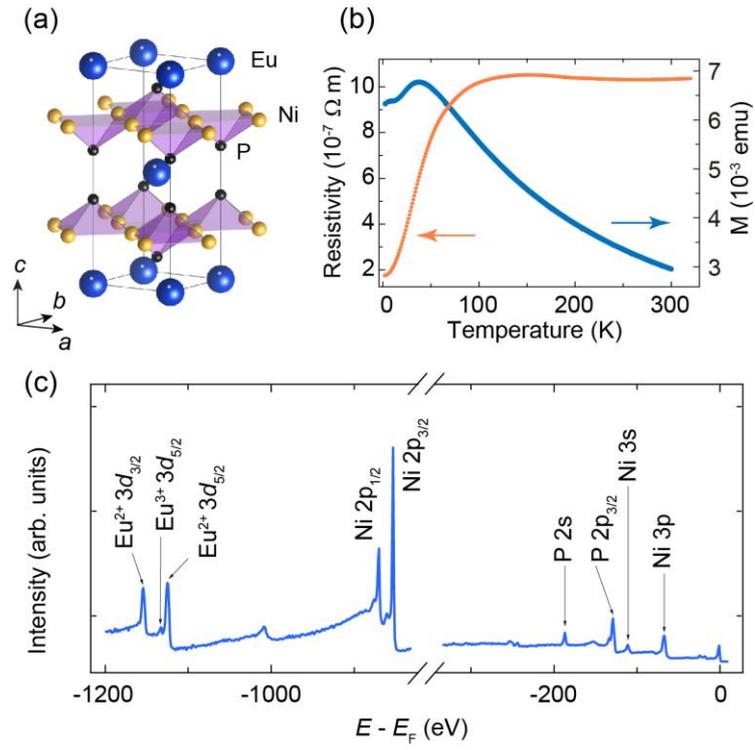

FIG. 1 (a) Schematic illustration of the crystal structure of $EuNi_2P_2$. (b) Resistivity and magnetization as a function of temperature with magnetic field applied along the $c$ axis. (c) X-ray photoemission spectrum showing the characteristic $Eu^{2+}$ and $Eu^{3+}$ $3d$-doublets as well as Ni and P core levels.

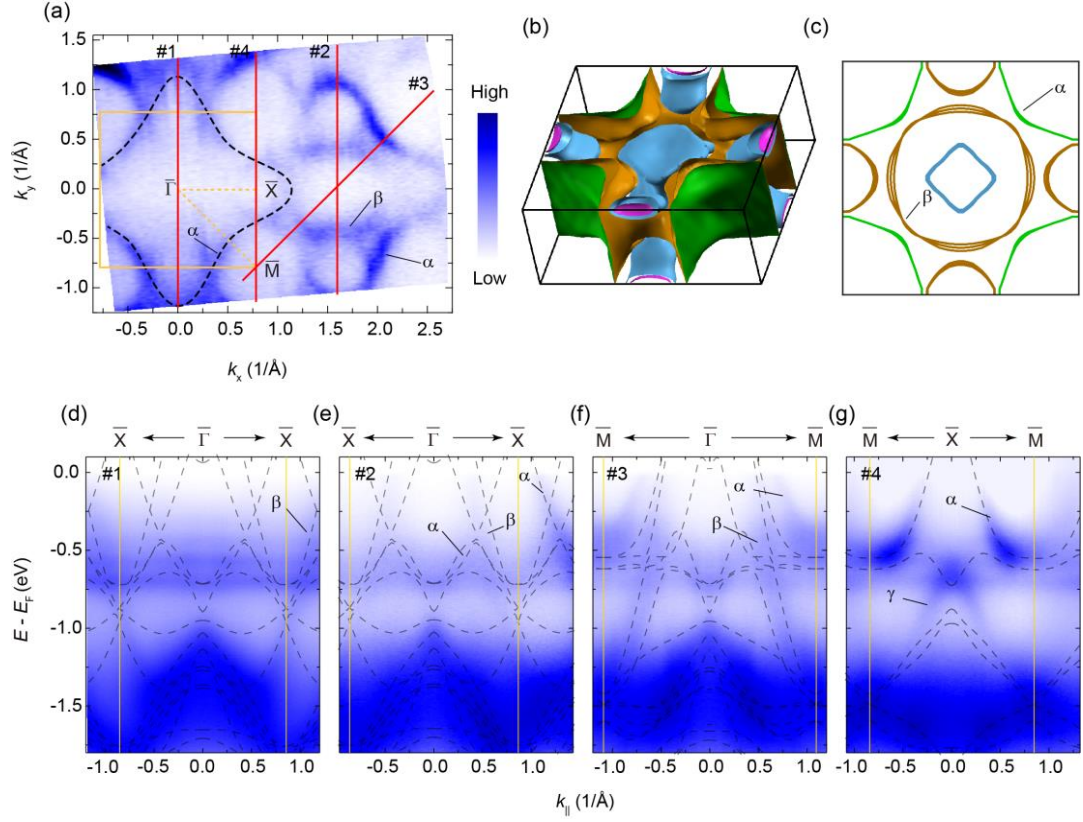

FIG. 2 (a) Fermi surface (FS) of EuNi$_2$P$_2$ measured by integrating ARPES intensity over an energy window of 30 meV around the Fermi level ($E_F$). (b), (c) Calculated FS in the three-dimensional Brillouin zone and the Fermi surface on the Γ$XM$ plane. (d)-(g) Band dispersions along high-symmetry directions as indicated in panel (a). The dashed lines are *ab-initio* calculations without *f* electrons but with spin-orbit coupling effect included. Data were collected under 105 eV photon energy at 15 K.

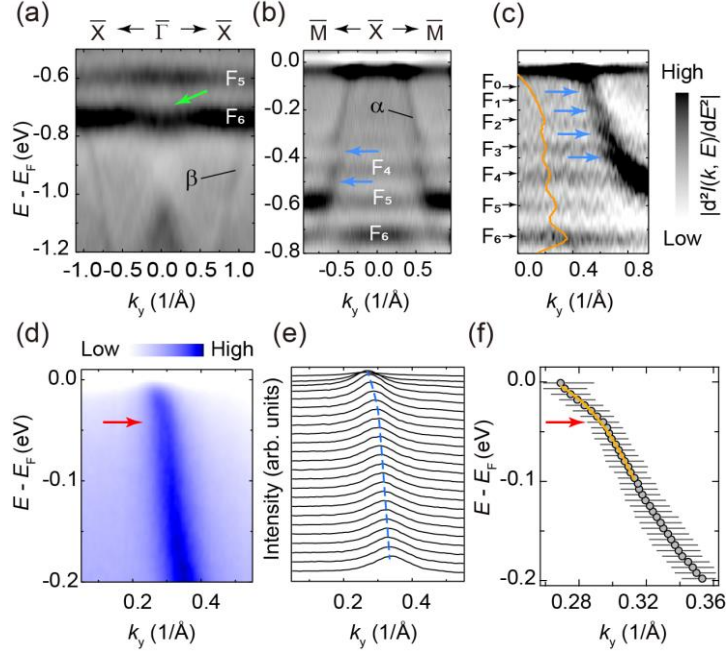

FIG. 3 (a)-(c) Second derivative of ARPES spectra along high-symmetry directions measured at (a) 105 eV, (b) 55 eV, and (c) 26 eV. The blue and green arrows indicate the hybridization induced dispersion anomalies. The orange curve is the integrated energy distribution curve (EDC) near Γ. (d) Zoom-in plot of ARPES spectra near $E_F$ along $\bar{X}\bar{M}$. (e) Stacking plot of the momentum distribution curves (MDCs) of the spectrum in (d). The blue dashed line is the guide to eye for the dispersion of the α band near $E_F$. (f) Band dispersion extracted from (d) by fitting the MDCs with a Lorentzian. The orange curve is the fit of the dispersion to the PAM. The red arrows in (d) and (f) indicate the kink in the dispersion near 40 meV below $E_F$.

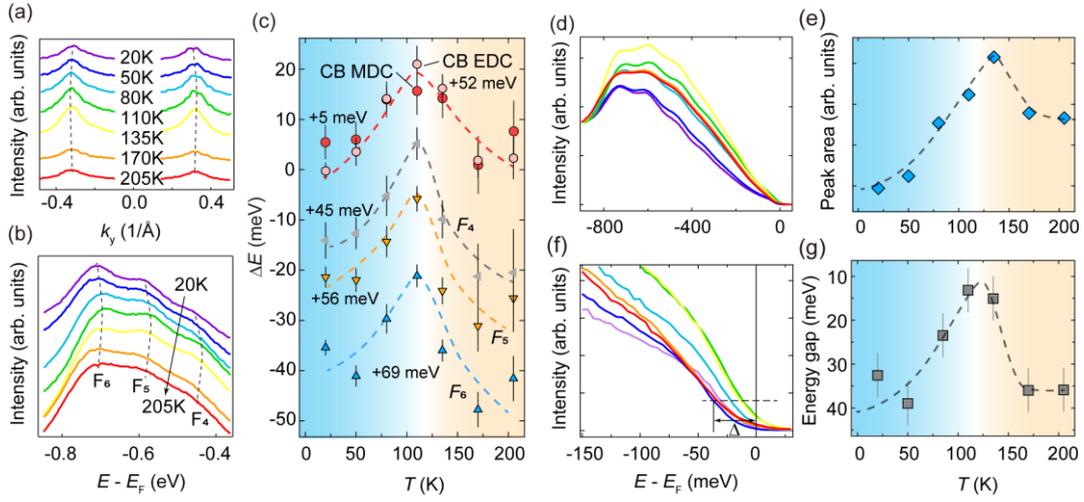

FIG. 4 (a) Temperature evolution of MDCs at $E_F$ along $\bar{X}\bar{M}$. (b) Temperature evolution of EDCs integrated between 0.4 and 0.7 Å$^{-1}$ (dominated by the $f$ bands) along $\bar{\Gamma}\bar{X}$. (c) Band shift as a function of temperature. Data in (a)-(c) were vertically shifted for clearness. CB EDC/MDC: conduction band shift obtained by fitting the EDCs/MDCs. $F_4$-$F_6$ are the representative $f$ bands. (d) The same as (b) but without vertical shift to show the change of the spectral weight. (e) Temperature evolution of the peak area in (d) integrated between -0.9 eV and $E_F$. (f) Zoom-in plot of the EDCs at the Fermi momentum of the α band showing a temperature-dependent energy gap near $E_F$. (g) Temperature dependence of the energy gap extracted by the "leading edge" position as shown in (f). The dashed lines in(a)-(c), (e), and (g) are guides to eyes.